\documentclass[a4paper,aps,prd,10pt,preprintnumbers,twocolumn,superscriptaddress,nofootinbib,amsmath,amssymb]{revtex4-1}
\usepackage{graphicx}
\usepackage[utf8]{inputenc}
\usepackage[T1]{fontenc}
\usepackage{cmap}

\def\imo{i}

\newcommand{\ie}{{i.e.,}~}
\newcommand{\eg}{{e.g.,}~}

\begin{document}
\title{General parametrization of black holes: the only parameters that matter}
\author{R. A. Konoplya} \email{roman.konoplya@gmail.com}
\affiliation{Institute of Physics and Research Centre of Theoretical Physics and Astrophysics, Faculty of Philosophy and Science, Silesian University in Opava, CZ-746 01 Opava, Czech Republic}
\affiliation{Peoples Friendship University of Russia (RUDN University), 6 Miklukho-Maklaya Street, Moscow 117198, Russian Federation}
\author{A. Zhidenko} \email{olexandr.zhydenko@ufabc.edu.br}
\affiliation{Centro de Matemática, Computação e Cognição (CMCC), Universidade Federal do ABC (UFABC),\\ Rua Abolição, CEP: 09210-180, Santo André, SP, Brazil}

\begin{abstract}
The general parametrization of a black-hole spacetime in arbitrary metric theories of gravity includes an infinite set of parameters.
It is natural to suppose that essential astrophysically observable quantities, such as quasinormal modes, parameters of shadow, electromagnetic radiation and accreting matter in the vicinity of a black hole, must depend mostly on a few of these parameters. Starting from the parametrization for spherically symmetric configurations in the form of infinite continued fraction, we suggest a compact representation of the asymptotically flat \emph{spherically symmetric} and \emph{slowly rotating} black holes in terms of only \emph{three} and \emph{four} parameters respectively. A subclass of arbitrarily rotating black holes belonging to the Carter family can also be parametrized by only four parameters. This approximate representation of a black-hole metric should allow one to describe physical observables in the region of strong gravity.
\end{abstract}
\maketitle

\section{Introduction}
Recent development of observations of black holes in gravitational and electromagnetic spectra \cite{alternative1}, despite current large uncertainties in measurement of black-hole parameters \cite{alternative2}, promises to determine the near-horizon geometry of black holes in the future and, thereby, to test the Einstein theory and its alternatives in the strong gravity regime. Therefore, it is important to have a general parametrized description of a black-hole spacetime in arbitrary metric theories of gravity, similar in the spirit to the parametrized post-Newtonian (PPN) formalism, and valid not only far from the black hole, but in the whole space outside the event horizon. Indeed, such a unified description allows one to consider various phenomena not in each theory of gravity, case by case, but using the general representation, so that constraining of the parameters there would show which theory of gravity is closer to the experimental data. For spherically symmetric black holes this parametrization was suggested in the form of the infinite continued fraction expansion in terms of the compact radial coordinate \cite{Rezzolla:2014mua}. It was further extended to the case of axially symmetric black holes \cite{Konoplya:2016jvv} and used for finding a number of analytical black-hole metrics \cite{Kokkotas:2017zwt,Kokkotas:2017ymc,Konoplya:2019goy,Hennigar:2016gkm,Konoplya:2019fpy} approximating numerical solutions \cite{EW,Kanti:1995vq,Antoniou:2017acq,Herdeiro:2018wub}. Various phenomena in the background of these parametrized black-hole metrics, such as quasinormal modes (QNMs) \cite{reviews}, particle motion, Hawking radiation \cite{Hawking:1974sw} and others, were studied in \cite{pappl}.

In the general case the parametrization includes an infinite number of parameters. However, it is natural to expect that physical quantities, which are potentially observable in astrophysical phenomena around black holes, must depend mostly on a few of these parameters. In addition, one would not believe that in the true theory of gravity these observable quantities deviate from their Schwarzschild values by orders, rather than by, at most, tens of percents. Otherwise, so strong deviations would be visible in the weak-field regime as well. The exception would occur supposing that the black-hole metric has the Kerr form in the whole space, except a very small region near its surface, where the deviation is strong. Then, such a geometry would be almost indistinguishable from the Kerr one, leaving a weak imprint only in the form of gravitational echoes at late times, when the signal is strongly damped \cite{Cardoso:2016rao}.

When considering a parametrized approximate representation of some exact black-hole solution one should  formulate the criterium of sufficient accuracy of the approximation. The physical ``effect'' which must be tested in the course of experiments is the deviation of one or another physical quantity (such as QNMs, parameters of the shadow, etc.) from their Schwarzschild values. Therefore it is natural to require that this effect must be at least one order larger than the relative error of the approximation due to the truncation of the infinite series.

In the present paper we consider a great number of examples of black-hole metrics and show that a spherically symmetric asymptotically flat black hole can be very well approximated by the following line element
\begin{eqnarray}\nonumber
ds^2&=&-N^2(r)dt^2+\frac{B^2(r)}{N^2(r)}dr^2+r^2 (d\theta^2+\sin^2\theta d\phi^2),\\\label{metric}
N^2(r)&=&1 - \frac{r_{0}(\epsilon+1)}{r} + \frac{r_{0}^3(\epsilon+a_1)}{r^3} - \frac{r_0^4 a_{1}}{r^4}, \\\nonumber
B^2(r)&=&\left(1+\frac{r_0^2 b_{1}}{r^2}\right)^2.
\end{eqnarray}
Here $r_0$ is the event horizon, so that $N(r_0)=0$; $\epsilon$, $a_1$ and $b_1$ are some parameters, such that when they all are equal to zero, the Schwarzschild limit is reproduced.  Within the approximation (\ref{metric}) the deviation of observable quantities are at least one order larger than the relative error. For more accurate approximation, such that the error is two orders smaller than the ``effect'', one can use a straightforward procedure to introduce additional coefficients, $a_2$ and $b_2$, in the metric functions. Further we show that this representation can be easily generalized to the case of slowly rotating black holes and mention some approaches to extension of this description to arbitrary rotation.

\section{The continued fraction expansion}
Following \cite{Rezzolla:2014mua}, we use the dimensionless variable $x \equiv 1-r_0/r$, so that $x=0$ corresponds to the event horizon, while $x=1$ corresponds to spatial infinity. In addition, we rewrite the metric function $N$ as $N^2=x A(x)$, where $A(x)>0$ for \mbox{$0\leq x\leq1$}.
Using the new parameters $\epsilon$, $a_0$, and $b_0$, the functions $A$ and $B$ can be written as
\begin{eqnarray}\nonumber
A(x)&=&1-\epsilon (1-x)+(a_0-\epsilon)(1-x)^2+{\tilde A}(x)(1-x)^3\,,
\\
B(x)&=&1+b_0(1-x)+{\tilde B}(x)(1-x)^2\,.
\end{eqnarray}
Here the coefficient $\epsilon$ measures the deviation of $r_0$ from $2M$,
$$\epsilon = \frac{2 M-r_0}{r_0}.$$
The coefficients $a_0$ and $b_0$ can be seen as combinations of the PPN parameters:
$$a_0=\frac{(\beta-\gamma)(1+\epsilon)^2}{2}, \qquad b_0=\frac{(\gamma-1)(1+\epsilon)}{2}.$$
Current observational constraints on the PPN parameters imply \mbox{$a_0 \sim b_0 \sim 10^{-4}$}.

The functions ${\tilde A}$ and ${\tilde B}$ are introduced through infinite continued fraction in order to describe the metric near the horizon (\ie for $x \simeq 0$),
\begin{equation}\label{ABdef}
{\tilde A}(x)=\frac{a_1}{\displaystyle 1+\frac{\displaystyle
    a_2x}{\displaystyle 1+\ldots}}, \qquad
{\tilde B}(x)=\frac{b_1}{\displaystyle 1+\frac{\displaystyle
    b_2x}{\displaystyle1+\ldots}},
\end{equation}
where $a_1, a_2,\ldots$ and $b_1, b_2,\ldots$ are dimensionless constants to be constrained from observations of phenomena near the event horizon. At the horizon only the first two terms of the expansions survive,
$
{\tilde A}(0)={a_1},~
{\tilde B}(0)={b_1},
$
which implies that near the horizon only the lower order terms of the expansions are important.

\section{Observable quantities}

Conditionally, we could divide physical effects characterizing black holes in the regime of strong gravity into two categories.
The first type of physical processes are almost completely determined by the \emph{near-horizon zone}, \eg thermodynamic properties, Hawking radiation\footnote{Though, the fraction of radiation which reaches the distant observer is corrected by the grey-body factors which depend on the effective potential barrier surrounding the black hole.} or gravitational echoes at very late times, which appear due to a strong modification of a black-hole metric in a small region near the horizon  \cite{Cardoso:2016rao}. Whatever important and intriguing, none of these effects are likely to be observed for astrophysical black holes in the nearest future. The second type of physical processes are related to ongoing observations in the electromagnetic and gravitational spectra. Their characteristics are determined by the black-hole geometry in the region around the peak of the effective potential. For example, the position of the innermost stable circular orbit (ISCO) of the Schwarzschild black hole ($x=2/3$) defines the region, essential for accretion, while the peak of the function $P(x)\equiv (1-x)^2xA(x)$ ($x=1/3$ for the Schwarzschild black hole) stipulates the region which is essential for the photon sphere and the position of the shadow cast by a black hole as well as for values of QN frequencies radiated by the black hole. This region, located at some distance from the black hole, but not much farther than ISCO, we shall call \emph{the radiation region}. In the extreme cases, which we do not consider here, \eg for an extremely rotating black hole, these two regions may be approaching each other. Here, we construct a compact and simple representation for the black-hole metric with the help of only a few parameters which would be effective when describing the second class of processes related to plausible astrophysical observations. We shall further call such metrics \emph{moderate} and discuss conditions for the black hole to have a moderate metric.

\begin{table*}
\begin{tabular}{|l|rrrr|rrrr|rrrr|}
  \hline
black hole& $R_{sh}$ &   effect &    $E_1$  &     $E_2$  & $\lambda$ &    effect &     $E_1$  &   $E_2$ & $\Omega_{ISCO}$ & effect &     $E_1$  &    $E_2$   \\
   \hline
Æther1    & $1.666$ &  $35.9\%$ &       $0$ &        $0$ & $1.14826$ & $198.3\%$ &        $0$ &        $0$ & $0.030101$ & $77.9\%$ &        $0$ &        $0$ \\
Æther2    & $2.043$ &  $21.4\%$ &       $0$ &        $0$ & $0.67377$ &  $75.1\%$ &        $0$ &        $0$ & $0.046342$ & $65.9\%$ &        $0$ &        $0$ \\
KS        & $2.149$ &  $17.3\%$ & $1.729\%$ & $0.1674\%$ & $0.58866$ &  $52.9\%$ &  $5.234\%$ & $0.5282\%$ & $0.117155$ & $13.1\%$ &  $4.474\%$ & $0.4327\%$ \\
HE        & $1.929$ &  $25.7\%$ & $2.871\%$ & $0.3659\%$ & $0.82911$ & $115.4\%$ &  $7.958\%$ & $0.4749\%$ & $0.158422$ & $16.4\%$ & $10.443\%$ & $3.7952\%$ \\
Hayward   & $3.972$ &  $52.9\%$ & $4.031\%$ & $3.3394\%$ & $0.20282$ &  $47.3\%$ &  $2.213\%$ & $2.6678\%$ & $0.092482$ & $32.0\%$ &  $9.583\%$ & $7.9520\%$ \\
Bronnikov & $3.687$ &  $41.9\%$ & $0.126\%$ & $0.0323\%$ & $0.18628$ &  $51.6\%$ &  $0.158\%$ & $0.1026\%$ & $0.120621$ & $11.4\%$ &  $0.291\%$ & $0.0657\%$ \\
Bardeen   & $3.247$ &  $25.0\%$ & $0.194\%$ & $0.1486\%$ & $0.23945$ &  $37.8\%$ &  $0.624\%$ & $0.5249\%$ & $0.121428$ & $10.8\%$ &  $0.405\%$ & $0.2966\%$ \\
EdM       & $3.266$ &  $25.7\%$ & $0.078\%$ & $0.0229\%$ & $0.24206$ &  $37.1\%$ &  $0.974\%$ & $0.1061\%$ & $0.138402$ &  $1.7\%$ &  $0.172\%$ & $0.0412\%$ \\
EsM       & $3.084$ &  $18.7\%$ & $0.582\%$ & $0.3303\%$ & $0.27603$ &  $28.3\%$ &  $3.120\%$ & $2.1431\%$ & $0.143746$ &  $5.6\%$ &  $1.694\%$ & $0.7214\%$ \\
E-Weyl    & $1.916$ &  $26.3\%$ & $0.664\%$ & $0.5862\%$ & $0.72329$ &  $87.9\%$ &  $0.905\%$ & $0.7578\%$ & $0.026784$ & $80.3\%$ & $35.057\%$ &$35.3715\%$ \\
CFM1      & $2.598$ &       $0$ &       $0$ &        $0$ & $0.54433$ &  $41.4\%$ &  $1.823\%$ & $0.1732\%$ & $0.136083$ &      $0$ &        $0$ &        $0$ \\
JP1       & $2.027$ &  $22.0\%$ &       $0$ &        $0$ & $0.42855$ &  $11.3\%$ &  $0.518\%$ & $0.0064\%$ & $0.138963$ &  $2.1\%$ &        $0$ &        $0$ \\
\hline
EdGB      & $2.700$ &   $3.9\%$ & $0.345\%$ & $0.2299\%$ & $0.36206$ &   $5.9\%$ &  $5.613\%$ & $1.3019\%$ & $0.131958$ &  $3.0\%$ &  $0.739\%$ & $0.6754\%$ \\
EsGB1     & $2.699$ &   $3.9\%$ & $0.386\%$ & $0.2206\%$ & $0.36245$ &   $5.8\%$ &  $5.494\%$ & $1.2332\%$ & $0.132027$ &  $3.0\%$ &  $0.900\%$ & $0.5882\%$ \\
EsGB2     & $2.868$ &  $10.4\%$ & $1.197\%$ & $1.0279\%$ & $0.32947$ &  $14.4\%$ & $12.916\%$ & $1.3233\%$ & $0.127488$ &  $6.3\%$ &  $2.786\%$ & $2.3443\%$ \\
CFM2      & $2.598$ &       $0$ &       $0$ &        $0$ & $0.31740$ &  $17.5\%$ & $56.759\%$ &  $4.330\%$ & $0.136083$ &      $0$ &        $0$ &        $0$ \\
JP2       & $2.270$ &  $12.6\%$ & $28.91\%$ & $9.4261\%$ & $0.43759$ &  $13.7\%$ &  $5.978\%$ & $14.963\%$ & $0.310425$ &$128.1\%$ & $87.588\%$ &$13.2753\%$ \\
  \hline
\end{tabular}
\caption{Radius of shadow, Lyapunov exponent and frequency at ISCO for a number of black holes, the relative effect compared to the Schwarzschild, and relative errors, $E_1$ and $E_2$, due to approximations of the first and second orders, respectively.}\label{tabl}
\end{table*}

For this purpose we consider the general parametrization \cite{Rezzolla:2014mua} and see at how many orders of the continued fraction expansion (\ref{ABdef}) this parametrization can be truncated in order to describe the above astrophysical processes for a great number of black-hole metrics. We shall measure all dimensional quantities in units of radius of the event horizon. The particular values of the parameters in the metrics under consideration are chosen in order to achieve considerable deviation of observables from the Schwazrschild black hole. However, this is not always possible, as in some cases, \eg in Gauss-Bonnet theories, the black-hole geometry reaches its extremal state already under a very small deviation from the Schwarzschild limit.
Our bunch of metrics are given in table~\ref{tabl} in abbreviated forms and includes: two particular examples of black holes in the Einstein-æther theory (Eq.~(51)~of~\cite{Berglund:2012bu} with $c_{13}r_{\mbox{\scriptsize æ}}^4=0.9$ (Æther1) and Eq.~(58)~of~\cite{Berglund:2012bu} with $r_u=0.9$ (Æther2), the quantum corrected black hole obtained by Kazakov and Solodukhin \cite{Kazakov:1993ha} for $a=0.86$ (KS), a number of regular black-hole metrics obtained in the context of non-linear electrodynamics or within other approaches (in the Heisenberg-Euler electrodynamics for $Q=0.1$ and $a = 10^4$ \cite{Yajima:2000kw} (HE), Hayward metric \cite{Hayward:2005gi} for $q=0.85$, Bronnikov metric \cite{Bronnikov:2000vy} for $M =0.95$, Bardeen spacetime \cite{Bardeen} for $a=0.5$). Various black holes with a scalar field and higher curvature corrections are considered: Einstein-dilaton-Maxwell black hole \cite{Garfinkle:1990qj} ($\phi_0=0$) for $Q=1$ (EdM), the black hole with a coupling $f(\phi)=\exp{(5\phi^2)}$ between Maxwell and scalar fields \cite{Herdeiro:2018wub,Konoplya:2019goy} for $P_0=0.55$ (EsM), black holes in theories with higher curvature corrections, such as Einstein-Weyl gravity \cite{Kokkotas:2017zwt,EW,Einstein-Weyl} for $p=1.1$ (E-Weyl), Einstein-dilaton-Gauss-Bonnet \cite{Kokkotas:2017ymc,Kanti:1995vq} for $p=0.6$ (EdGB) and its generalization to other couplings to the scalar field, Einstein-scalar-Gauss-Bonnet \cite{Antoniou:2017acq,Konoplya:2019fpy}, with the coupling $f(\phi)=1/(4\phi)$ for $p=0.6$ (EsGB1) and the coupling $f(\phi)=\ln(\phi)/4$ for $p=0.99$ (EsGB2). In addition, several other examples were studied: Casadio-Fabbri-Mazzacurati black hole \cite{Casadio:2001jg,Germani:2001du} for $r_0=0$ (CFM1) and $r_0/(2M)=0.99$ (CFM2) in the context of the brane-world model, and Johannsen-Psaltis ad hoc metrics \cite{Johannsen:2011dh} for the nonzero values of $\epsilon_2 = 5$ and $\epsilon_3 = 0.5$ (JP1) and for the only nonzero $\epsilon_{10} = 4 \cdot 10^{3}$ (JP2).

The simple and illustrative characteristics, which we consider here, are: radius of the black-hole shadow $R_s$, which can be found if one calculates the maximum of the function $P(x)$, \begin{equation}
\frac{r_0^2}{R_{s}^2}=\max P(x)=P(x_m),
\end{equation}
and the Lyapunov exponent $\lambda$, which depends on the second derivative in the same point $x_m$,
\begin{equation}
\lambda^2 = -\frac{P(x_m)P''(x_m)}{2r_0^2B^2(x_m)}.
\end{equation}
Eikonal regime of QNMs \cite{eikonal} can be represented through the frequency of the null geodesics and the Lyapunov exponent \cite{Cardoso:2008bp,Khanna:2016yow,Konoplya:2017wot,Jusufi:2019ltj}. Thus, one can easily deduce
\begin{equation}
\omega=\frac{1}{R_{s}}\left(\ell+\frac{1}{2}\right)-\imo\lambda\left(n+\frac{1}{2}\right)+{\cal O}\left(\frac{1}{\ell}\right).
\end{equation}

The basic quantity characterizing the effectiveness of the truncation of the continued fraction at a given order is the ratio of the ``effect'', that is of the deviation of the observable quantity from its Schwarzschild limit for the exact metric, to the relative error due to truncation of its analytic approximation.

The top part of table~\ref{tabl} contains examples of black holes with moderate metrics.
From the data shown in table~\ref{tabl} we see that the simplest Æther1 and Æther2\footnote{Æther2 black hole has an effective Reissner-Nordström line element, so that $a_1=b_1=0$.}
metrics are reproduced exactly already at the first order of approximation, representing a trivial example. For other examples the truncation of the continued fraction at the first order provides the relative error, which is at least one order less than the effect. Second-order approximation further increases the accuracy. For sufficiently small deviations from Schwarzschild, each order of the approximation increases accuracy at least by one order \cite{Rezzolla:2014mua}. However, when the relative effect is quite large (as in table~\ref{tabl}), the second-order approximation improves the accuracy by tens of percents or more.

The bottom part of table~\ref{tabl} is devoted to the examples, for which the error of the first-order approximation is comparable with the effect. We notice that even for near-extremal cases the effect for shadows and Lyapunov exponents (hence for QNMs) are relatively small and hardly exceed $10 \%$ . For the EdGB theory the black-hole geometry deviates only slightly from the Einstein limit, so that even the near extreme black holes allows for effect of only a few percents \cite{Younsi:2016azx,Cunha:2016wzk,Nampalliwar:2018iru}. This is in the agreement with our intuitive definition of such geometries, which deviate from the Schwarzschild black hole only near the horizon.

In order to estimate the accurateness of the approximation in the radiation region we also calculate radius of the innermost stable circular orbit (ISCO) of a massive particle. In table~\ref{tabl} we present the orbital frequency at ISCO, $\Omega_{ISCO}$, which is an observable quantity, and the relative errors. Again we see that for the moderate metrics the error is considerably smaller than the effect already at the first order of the expansion and the second-order approximation significantly improves the accuracy. The only exception is the E-Weyl black hole, which requires further increasing of the approximation orders for the chosen parameter $p=1.1$. However, such a black hole has a vanishing effective mass, being, thereby, not an appropriate object for a viable accretion process.

It is important to understand that relatively large error for the examples from the bottom of table~\ref{tabl} is not for the whole range of physical parameters within the corresponding theories. It usually occurs to near extremal black-hole states, while far from it our approximation still shows a good convergence.

The first-order approximation provides small relative error if the functions ${\tilde A}(x)$ and ${\tilde B}(x)$ (\ref{ABdef}) are well approximated by a constant, which is their value at the horizon, ${\tilde A}(0)=a_1$ and ${\tilde B}(x)=b_1$. The approximation by constant is good only if the functions $A(x)$ and $B(x)$ \emph{do not change strongly between the horizon and the radiation region.} Indeed, for the moderate metrics we observe the relatively slow change of the metric function $A(x)$ and $B(x)$, starting from the event horizon and until the radiation zone.

\begin{figure}
\centerline{\resizebox{\linewidth}{!}{\includegraphics*{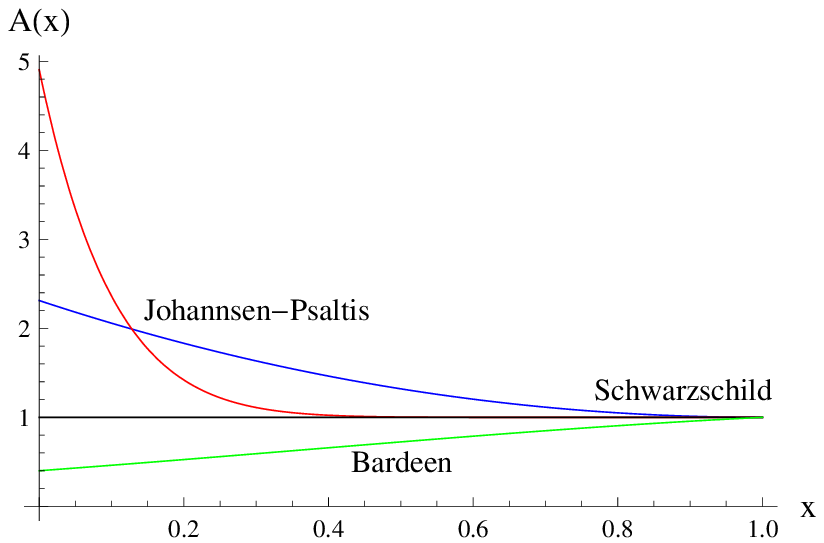}}}
\centerline{\resizebox{\linewidth}{!}{\includegraphics*{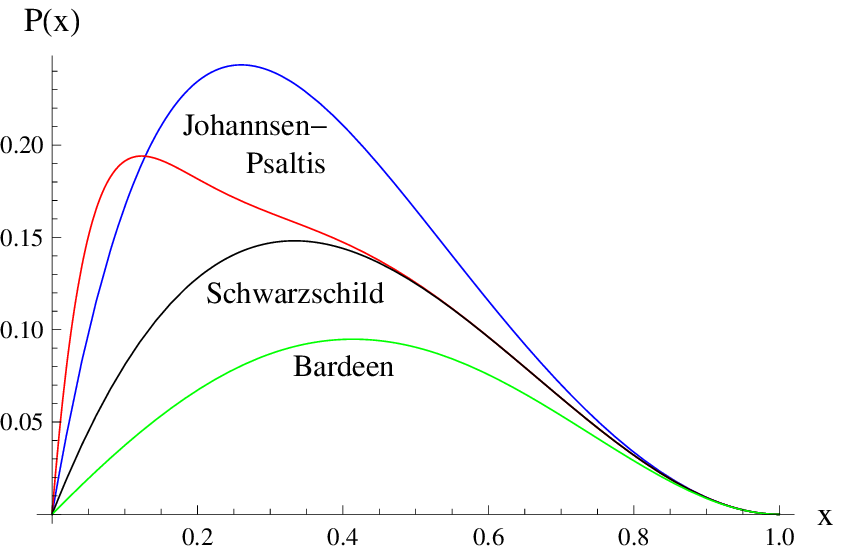}}}
\caption{Plots of $A(x)$ (top panel) and $P(x)\equiv (1-x)^2xA(x)$ (bottom panel) for JP2 black hole  (red, top), JP1 black hole (blue), Schwarzschild black hole (black), and Bardeen black hole (green, bottom).}\label{fig1}
\end{figure}

We present an illustration on Fig.~\ref{fig1} where we compare moderate metrics, JP1 and Bardeen, with rather an artificial example JP2 with vanishing all the lower  $\epsilon_i$ ($i<10$), except for an extremely high $\epsilon_{10}$. This example is designed to understand the cases which cannot be effectively described by our approach. The JP1 and Bardeen metrics are very well approximated already at the first order (see table~\ref{tabl}). Thus, only three parameters $a_1$, $b_1$ and $\epsilon$ are sufficient to describe the observable quantities. JP2 geometry, however, has the metric functions which change strongly near the event horizon and reaches its asymptotic regime at a relatively short distance from the black hole. This is appropriate to spacetimes representing, for example, various modifications of the black-hole geometry in the near-horizon region only due to quantum corrections or other new physics near the surface of the compact object. Such spacetimes are not likely to be distinguished from the Einsteinian black holes in the nearest future, because their geometries can be tested either via direct observation of Hawking radiation or, still elusive, echoes at very late times.

Thus, we conclude that for the astrophysically relevant observations (shadows, QNMs, accretion) the black hole with moderate metrics are the most important targets. Although for a stress test of our approach we considered here general post-Newtonian behavior, one can practically neglect the black-hole charge and choose $a_0=b_0=0$. The line element of such black holes can be well approximated by~(\ref{metric}).

In this case the problem of computation of $x_m$ is reduced to the solution of the quadric equation, hence the shadow radius ($R_s$) can be found in a closed but cumbersome form. It depends almost linearly on $a_1$ and decreases as $a_1$ grows,
\begin{eqnarray}\label{shadow-parametric-a}
  \frac{r_0^2}{R_{s}^2}&=&\frac{(1-x_0)^2x_0^2(2x_0-3)}{5 x_0^2 - 10 x_0 + 2}+(1-x_0)^5x_0 a_1\\\nonumber
  &+&\frac{(1-6x_0)^2(1-x_0)^8(5 x_0^2 - 10 x_0 + 2)}{12 (5 x_0^3 - 10 x_0^2 + 5 x_0 -1)}a_1^2+{\cal O}(a_1^3),
\end{eqnarray}
where $x_0$ is the compact coordinate for the photon circular orbit, satisfying the cubic equation,
\begin{equation}
1 - 2 \epsilon - 3(1 - 4 \epsilon)x_0 - 15 \epsilon x_0^2 + 5 \epsilon x_0^3 =0,
\end{equation}
and monotonously increases with $\epsilon$. For small $\epsilon$ we find
\begin{equation}
x_0=\frac{1}{3} + \frac{14}{81}\epsilon + \frac{154}{729}\epsilon^2 + \frac{3122}{19683}\epsilon^3 + {\cal O}(\epsilon^4).
\end{equation}
Similarly, for the Lyapunov exponent ($\lambda$) we obtain
\begin{eqnarray}\label{Lyapunov-parametric-a}
&&\lambda^2 R_s^2=\frac{3 (5 x_0^3 - 10 x_0^2 + 5 x_0 - 1)}{(5 x_0^2 - 10 x_0 + 2)^2(1+b_1(1-x_0)^2)^2}
\\\nonumber
&&-\frac{(1 + x_0)^5 (120 x_0^4 - 255 x_0^3 + 145 x_0^2 - 47 x_0 + 7)b_1a_1}{2 (5 x_0^3 - 10 x_0^2 + 5 x_0 - 1)(1+b_1(1-x_0)^2)^3}
\\\nonumber
&&-\frac{(1 + x_0)^3 (5 x_0^3 - 15 x_0^2 - 3 x_0 + 3)a_1}{2 (5 x_0^3 - 10 x_0^2 + 5 x_0 - 1)(1+b_1(1-x_0)^2)^3}+{\cal O}(a_1^2).
\end{eqnarray}
Since both quantities depend almost linearly on $a_1$, one can expect that the error due to the approximation remains one order smaller than the effect as long as the metric stays moderate. If one needs to achieve the approximation in which the error would be two orders less than the effect, then the second order can be used via consideration of non-zero $a_2$ and $b_2$ in (\ref{ABdef}).

The approximation (\ref{metric}) can be extended to the small rotation regime as
$$ds_a^2=ds^2-\frac{4Ma\sin^2\theta}{r}dtd\phi,$$
which implies that corrections owing to the modification of gravity must be much larger than those due to rotation, \ie $a/M\ll a_1,b_1$, but also that the second order corrections given by $a_2$ and $b_2$ are negligible. Thus, in the hierarchy of corrections, the above $\sim d t d \phi$-term is between the first- and second-order corrections in the radial direction.

In the case of generic rotation, and no further assumptions about the black-hole spacetime, we do not see an easy way to make the approximation as simple as in the spherical case.
However, for metrics allowing for separation of variables in the Klein-Gordon and Hamilton-Jacoby equations \cite{Carter:1968ks,Chen:2019jbs}, following \cite{Konoplya:2018arm} we can generalize (\ref{metric}) to the arbitrary rotation as follows
\begin{eqnarray}
ds^2&=&-\dfrac{N^2(r)-W^2(r,\theta)\sin^2\theta}{K^2(r,\theta)}dt^2
\\\nonumber
&-&2W(r,\theta)r\sin^2\theta dt \, d\phi+K^2(r,\theta)r^2\sin^2\theta d\phi^2 
\\\nonumber
&+&\Sigma(r,\theta)\left(\dfrac{B^2(r)}{N^2(r)}dr^2 +r^2d\theta^2\right),
\\\nonumber 
N^2(r)&=&\left(\!1+\dfrac{a^2}{r^2}\!\right)\left(\!1-\dfrac{r_0}{r}\!\right) - \dfrac{r_0\epsilon}{r} + \dfrac{r_0^3(\epsilon+a_1)}{r^3} - \dfrac{r_0^4a_1}{r^4}, 
\\\nonumber B^2(r)&=&\left(1+\frac{r_0^2 b_{1}}{r^2}\right)^2,
\end{eqnarray}
\begin{eqnarray}\nonumber
\Sigma(r,\theta)&=&1+\frac{a^2\cos^2\theta}{r^2}, 
\\\nonumber
W(r,\theta)&=&\frac{a(r^2+a^2-r^2N^2(r))}{r^3\Sigma(r,\theta)}, 
\\\nonumber
K^2(r,\theta)&=&\dfrac{r^2+a^2+a^2\cos^2\theta N^2(r)}{r^2\Sigma(r,\theta)}+\dfrac{a}{r}W(r,\theta).
\label{axisymmetric}
\end{eqnarray}
When $\epsilon=a^2/r_0^2$, $a_1=b_1=0$ we obtain the Kerr solution. The above parametrization is useful even for some metrics not allowing for the separation of variables (\eg for EdGB black hole) \cite{Konoplya:2018arm}. In order to test viability of the latter parametrization as thoroughly as the spherically symmetric one, we need many more examples of axially-symmetric black-hole solutions (analytical or numerical) in various theories of gravity, which are currently lacking.

\section{Conclusions}

We have shown that, in order to estimate essential astrophysical observable quantities of general spherically symmetric and slowly rotating black holes, as well as of arbitrarily rotating black holes belonging to a Carter subclass, it is sufficient to parameterize the spacetime of the black hole with only three parameters for the spherical case and four parameters for the rotating one. This compact form can considerably simplify further modelling of astrophysical phenomena in the background of a generic black hole and should help to constrain the black-hole geometry in the future. Compact objects which are characterized by a sudden change of the metric functions in the near-horizon zone only, as, for example, various black-hole mimickers similar to Damour-Solodukhin wormholes \cite{Damour:2007ap} or black holes with quantum corrections owing to the cloud of quantized fields in the Plank scale region near the event horizon, cannot be well approximated in this way, and, at the same time, are not likely to affect characteristics of radiation in the gravitational and electromagnetic spectra (such as shadows, QNMs and accretion) considerably.

\acknowledgments
R.~A.~K. thanks 19-03950S GAČR grant and ``RUDN University Program 5-100''.
A.~Z. was supported by Conselho Nacional de Desenvolvimento Científico e Tecnológico (CNPq).

\end{document}